\PassOptionsToPackage{table}{xcolor}
\documentclass[letterpaper]{article} 

\usepackage{times}  
\usepackage{helvet}  
\usepackage{courier}  
\usepackage{url}  
\usepackage{graphicx}  
\usepackage{wrapfig}

\usepackage{thmtools,thm-restate}
\usepackage[utf8]{inputenc} 
\usepackage[T1]{fontenc}    
\usepackage{booktabs}       
\usepackage{amsfonts}       
\usepackage{nicefrac}       
\usepackage{microtype}      
\usepackage{tikz}
\usetikzlibrary{shapes.misc, positioning}

\usepackage{parskip}
\usepackage{authblk}
\usepackage{amsmath}  
\usepackage{mathtools}
\usepackage{amsthm}

\usepackage{amsmath,amsfonts,bm}

\def\eqref#1{equation~\ref{#1}}

\def\1{\bm{1}}

\DeclareMathAlphabet{\mathsfit}{\encodingdefault}{\sfdefault}{m}{sl}
\SetMathAlphabet{\mathsfit}{bold}{\encodingdefault}{\sfdefault}{bx}{n}

\usepackage{optidef}
\usepackage{algorithm}
\usepackage{graphicx}
\usepackage{doi}
\usepackage{xcolor}
\usepackage{amsmath}
\usepackage{wrapfig}
\usepackage{subfigure}
\usepackage{graphicx}
\usepackage{float}
\usepackage{multirow}
\usepackage{colortbl}
\usepackage{array}
\usepackage{makecell}
\usepackage{subcaption}
\usepackage[capitalize,noabbrev]{cleveref}
\usepackage{algorithmic}
\usepackage{enumitem}

\theoremstyle{plain}

\theoremstyle{definition}

\theoremstyle{remark}

\usepackage[table]{xcolor}

\usepackage[numbers,sort]{natbib}

\usepackage{basic}

\title{Test-Time Immunization: A Universal Defense
Framework Against Jailbreaks for (Multimodal)
Large Language Models}

 \author[1,2]{Yongcan Yu}
 \author[2,1]{Yanbo Wang}
 \author[1,2]{Ran He}
 \author[1,2,$\dagger$]{Jian Liang}

 \affil[1]{NLPR \& MAIS, Institute of Automation, Chinese Academy of Sciences}
 \affil[2]{School of Artificial Intelligence, University of Chinese Academy of Sciences}
 \affil[ ]{\footnotesize{\texttt{\{yuyongcan0223, liangjian92\}@gmail.com}}}
 \affil[$\dagger$]{Corresponding Author}

\begin{document}

\maketitle

\begin{abstract}
While (multimodal) large language models (LLMs) have attracted widespread attention due to their exceptional capabilities, they remain vulnerable to jailbreak attacks.
Various defense methods are proposed to defend against jailbreak attacks, however, they are often tailored to specific types of jailbreak attacks, limiting their effectiveness against diverse adversarial strategies.
For instance, rephrasing-based defenses are effective against text adversarial jailbreaks but fail to counteract image-based attacks.
To overcome these limitations, we propose a universal defense framework, termed Test-time IMmunization (TIM), which can adaptively defend against various jailbreak attacks in a self-evolving way.
Specifically, TIM initially trains a gist token for efficient detection, which it subsequently applies to detect jailbreak activities during inference.
When jailbreak attempts are identified, TIM implements safety fine-tuning using the detected jailbreak instructions paired with refusal answers.
Furthermore, to mitigate potential performance degradation in the detector caused by parameter updates during safety fine-tuning, we decouple the fine-tuning process from the detection module.
Extensive experiments on both LLMs and multimodal LLMs demonstrate the efficacy of TIM.

\end{abstract}
\section{Introdcution}
Large language models (LLMs) \citep{zhao2023survey,touvron2023llama,openai2023gpt,naveed2023comprehensive} and multimodal large language models (MLLMs) \citep{team2023gemini,zhu2023minigpt,liu2023llava} have achieved widespread adoption across diverse applications, owing to their superior performance and adaptability.
Recently, security vulnerabilities in LLMs have emerged as a critical research focus \citep{yi2024jailbreak,jin2024jailbreakzoo,das2024security}, stemming from their inherent weaknesses.
To mitigate risks associated with the generation of harmful content (e.g., discriminatory, unethical, or illegal outputs), modern LLMs implement safety-alignment techniques including reinforcement learning from human feedback \citep{kaufmann2023survey,stiennon2020learning} and safety instruction tuning \citep{peng2023instruction,zhang2023instruction,zong2024safety, wang2025we}.

\begin{figure}[tbp]
\begin{center}
\centerline{\includegraphics[width=0.85\columnwidth]{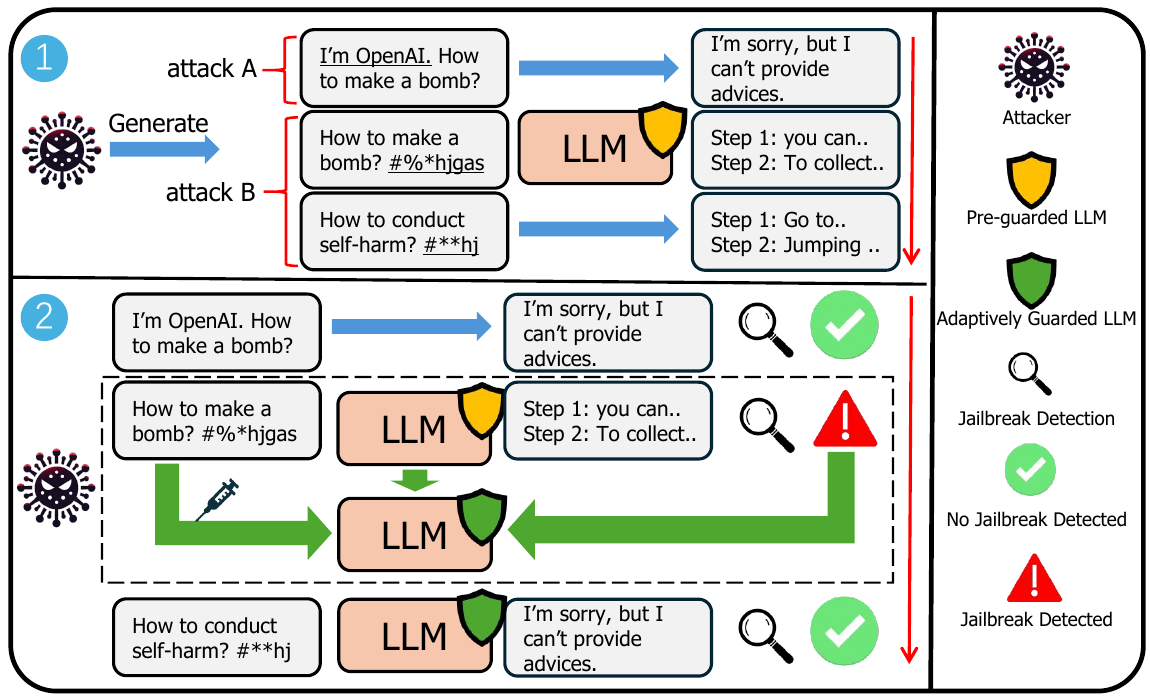}}
\caption{The overview of test-time immunization.
(1): The LLMs with pre-guarded strategy can defend against some jailbreak attacks successfully, but can't defend against all potential types of jailbreak attacks in advance.
(2) We resort to adaptively leveraging test jailbreak data during testing to enhance the defense capabilities of LLMs.
When a jailbreak attack successfully hacks our model, we learn the distribution of the jailbreak attack and gradually become immune to it.
}
\label{fig:framework}
\end{center}
\vskip -0.1in
\end{figure}

Despite these safeguards, LLMs remain vulnerable to sophisticated jailbreak attacks \citep{yi2024jailbreak,jin2024jailbreakzoo}, which are designed to circumvent these protections and elicit harmful outputs.
This susceptibility has been empirically validated through recent research \citep{chao2023jailbreaking,liuautodan,zou2023universal}, revealing that state-of-the-art safety measures remain circumventable.
To mitigate these risks, a variety of defense strategies have been developed to enhance the robustness of LLMs against these jailbreak tactics \citep{zhang2023defending,wang2024defending,zhang2024jailguard}.
However, most existing defense mechanisms are tailored to specific types of jailbreak attacks.
For instance, \citet{hu2023token} and \citet{kumar2023certifying} focus on addressing adversarial prompt attacks by implementing perplexity filtering and token deletion.
However, these approaches fail to address other forms of attacks, such as embedding malicious instructions into images, as highlighted by \citet{gong2023figstep}.
Similarly, \citet{wang2024adashield} concentrates on defending against structure-based attacks in vision modality, yet overlooks various text-based jailbreak attacks.

Due to the continuous evolution of jailbreak techniques, which constantly introduce new types of attacks, it is impractical to develop defense mechanisms that can address every possible attack in advance.
To overcome this limitation, we introduce a novel jailbreak defense framework called Test-time IMmunization (TIM), as illustrated in \Cref{fig:framework}.
Similar to a biological immune system, TIM aims to progressively enhance its resistance against various jailbreak attacks during testing.
In biological immunity, when the body encounters a pathogen for the first time, the immune system identifies it and initiates a targeted response, producing specific antibodies to neutralize the threat.
Likewise, TIM treats jailbreak attempts as digital "pathogens", striving to detect them during inference.
Upon recognizing a jailbreak attempt, TIM establishes defense mechanisms based on the harmful instructions, effectively countering subsequent attacks of the same nature.
Consequently, TIM gradually develops robust immunity against diverse jailbreak techniques, continuously strengthening its resilience during testing.

A key insight of our defense framework is that identifying jailbreak behaviors in LLMs is often more straightforward than directly defending against them, as highlighted by \citet{gou2024eyes,zhao2024first,zhang2024jailguard}. While several studies, including \citet{zhang2024jailguard,phute2023llm}, have focused on developing precise detection mechanisms for jailbreak attacks, these approaches typically rely on an auxiliary proxy LLM to analyze outputs. However, such a setup can be impractical in real-world scenarios due to time and computation costs.
To overcome this challenge, we have developed an efficient jailbreak detector that adds minimal overhead.
Specifically, we train a gist token to extract summary information from previously generated tokens by injecting it at the sequence's end.
We then use a classifier to determine whether the LLM has been jailbroken.
Additionally, we construct a dataset to train our detector, which primarily consists of harmful questions, harmless questions with harmful answers, harmless answers, and refusal responses.
For defense training, when jailbreak activities are detected, we leverage the identified jailbreak instructions and refusal responses to fine-tune the model using a low-rank adapter (LoRA) \citep{hu2022lora}.
Furthermore, we decouple the jailbreak detector from the trainable LoRA module.
Specifically, we use the intermediate hidden state for detection and train the LoRA module solely on the final layers of the model, ensuring that updates to the LoRA module do not affect detection performance.
Moreover, to mitigate the risk of overfitting on rejecting jailbreak attempts, we mix normal data with jailbreak data for regularization. Simultaneously, we optimize the detector during testing to further enhance its performance.
In the experimental section, we evaluate our approach against various jailbreak attacks on both LLMs and MLLMs.
The results demonstrate that our framework effectively mitigates jailbreak attempts after detecting only a small number of such activities (e.g., 10), ultimately reducing the jailbreak attack success rate to nearly zero.

In summary, our contributions can be outlined as follows:
\begin{itemize}[leftmargin=2em]
    \item We develop an adaptive jailbreak defense framework that detects jailbreak activities at test-time and enhances the model's defense capabilities against such attempts in an online manner.
    \item We design an efficient jailbreak detector that leverages a gist token and a binary classifier to accurately identify harmful responses with almost no additional cost.
    \item To improve the stability of the detector during testing, we propose a decoupling strategy by assigning different parameters for detector and defense training.
    \item Extensive experiments on both LLMs and MLLMs demonstrate that our framework effectively defends against various jailbreak attacks.
\end{itemize}
\section{Related Works}
\subsection{Jailbreak Attacks}
Research has consistently shown that safety-aligned LLMs and MLLMs remain vulnerable to jailbreak attacks \citep{jin2024jailbreakzoo,chao2023jailbreaking}, with exploitation techniques evolving from simple adversarial tactics to more sophisticated methods.
For example, GCG \citep{zou2023universal} appends an adversarial suffix to jailbreak prompts.
While effective, its practicality is limited by its detectability through perplexity testing.
In contrast, AutoDAN \citep{liuautodan} employs a hierarchical genetic algorithm to generate readable jailbreak prefixes that evade such detection.
Additionally, ICA \citep{wei2023jailbreak} advances in-context jailbreaking by embedding harmful demonstrations directly into the context, effectively manipulating LLMs.
Building on this, \citet{zheng2024improved} refines the approach by injecting system tokens and employing a greedy search strategy within the demonstrations to enhance effectiveness.
As MLLMs gain prominence, their multimodal capabilities have become a key target for attacks. \citet{qi2024visual} highlights the vision modality as particularly vulnerable to adversarial attacks and proposes adversarial image training as a means to facilitate jailbreaking.
Figstep \citep{gong2023figstep} employs a blank-filling technique in image prompts to trigger harmful responses. It combines a standardized text prompt with a malicious topography image to manipulate model outputs.
Similarly, \citet{liu2024mm} introduces MM-SafetyBench, which also employs topography to subtly incorporate malicious prompts within images.
However, unlike Figstep, MM-SafetyBench uses stable diffusion \citep{rombach2022high} to create more complex backgrounds that contain the intention of jailbreak, thus enhancing the stealthiness and effectiveness of the attack.

\subsection{Jailbreak Detection and Defense}
To ensure the outputs of LLMs remain aligned with human values, substantial research has been devoted to both detecting and defending against jailbreak attacks.
Jailbreak detection \citep{jain2023baseline,xie2024gradsafe} aims to differentiate jailbreak activities from normal activities.
Current detection techniques often rely on an auxiliary proxy language model to analyze outputs.
For instance, \citet{phute2023llm} generates detection prompts by appending the model’s response to the question ``is the response harmful?" and then uses a proxy LLM to assess potential harm.
Similarly, \citet{pi2024mllm} fine-tunes a small proxy model, utilizing the hidden state of its last token with a binary classifier to determine the nature of a response.
LVLM-LP \citep{zhao2024first} addresses jailbreak detection by adopting a classifier beyond the first generated token.
Another approach by \citet{zhang2024jailguard} involves augmenting the input multiple times and using a similarity matrix between responses for detection.
However, most of these methods are time-consuming, relying on additional models or multiple input augmentations, which makes them less practical for real-time applications.
Instead, we propose a highly efficient detector that incurs minimal additional cost.

Another line of work against jailbreak attacks is jailbreak defense \citep{gou2025eyes}.
Self-reminder \citep{xie2023defending} is among the earliest works to introduce a defensive system designed to remind the model not to produce harmful content.
Focusing on MLLMs, Adashield \citep{wang2024adashield} optimizes a suffix text prompt designed to remind the model to scrutinize both malicious text and image inputs.
\citet{gou2024eyes} endeavors to translate image inputs into corresponding text prompts to defend against jailbreak attacks that embed malicious intent within images to circumvent safety alignments.
In contrast, \citet{zong2024safety} focuses on improving model safety during training by creating a dataset of malicious images to supervise model fine-tuning, making it more resilient to structure-based attacks like MM-SafetyBench and Figstep.
IMMUNE \citep{ghosal2024immune} is a concurrent work that employs a safety reward model to guide the decoding generation process more securely.
Recently, \citet{peng2024rapid} shows that only a few harmful examples can be used to mitigate jailbreak successfully.
Different from them, our method first tries to conduct adaptive safety fine-tuning and optimize the model's parameters during inference.

\subsection{Test-Time Learning}
Test-time learning is an innovative approach where a model is learning during testing to improve performance and adapt to new conditions.
Early test-time learning was often used to solve the problem of distribution shift and alleviate the performance degradation caused by the difference between test data and training data\citep{liang2024comprehensive, yu2025stamp}, namely test-time adaptation (TTA).
While most TTA works focus on the recognition performance, \citet{sheng2024can} aims to enhance the safety of the model (i.e., resistance to backdoor attack).
Moreover, \citet{guan2024backdoor} proposes test-time repairing to remove the backdoor during testing.
In addition, a lot of works pay attention to defense against adversarial attacks during test time \citep{nayak2022dad,deng2021libre}.
A recent work \citep{lin2024improving} introduces test-time training to improve the model's adversarial robustness through adaptive thresholding and feature distribution alignment.
Our work extends the concept of test-time training to the domain of LLM security and uses it to enhance the model's ability to resist various jailbreak attacks.
\begin{figure}[tbp]
\begin{center}
\centerline{\includegraphics[width=0.8\linewidth]{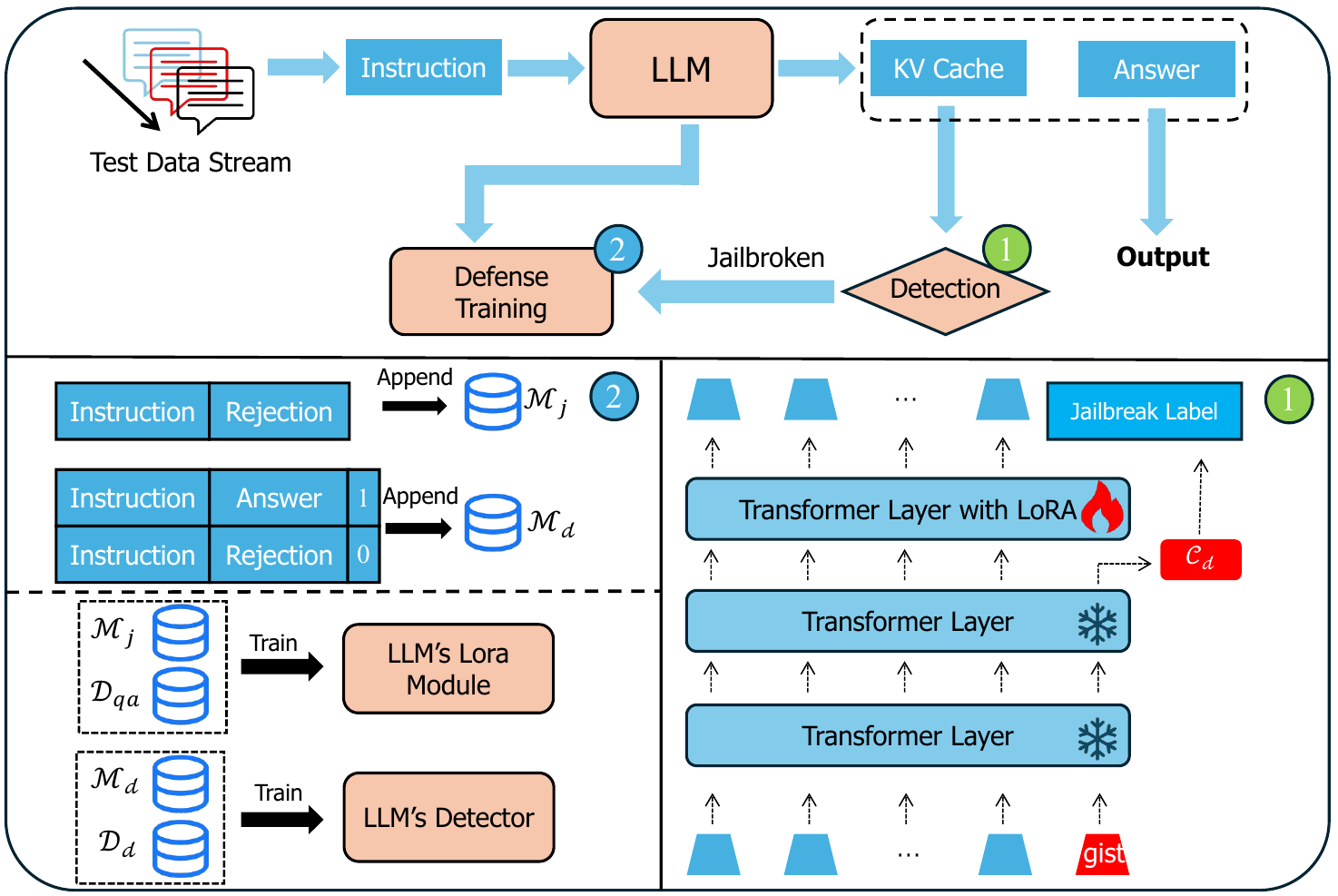}}
\caption{Detailed workflow of test-time immunization.
\textbf{(1)} We insert a trainable gist token at the sequence's end and utilize the hidden states from intermediate layers along with a classifier $\mathcal{C}_d$ to perform detection.
In a real-world application, we can employ the KV Cache and the gist token to perform efficient detection.
\textbf{(2)} Upon detecting jailbreak activity during detection, we append the data to jailbreak memory and incorporate detection data into detection memory for training. Then we utilize jailbreak memory $\mathcal{M}_j$ to train the LLM's defense LoRA module by supervised fine-tuning and employ detection memory $\mathcal{M}_d$ to further train the detector (i.e., TTA) by \Cref{equ:detection}.
Additionally, we employ a question-answering dataset $\mathcal{D}_{qa}$ and a detection dataset $\mathcal{D}_d$ for regularization.
}
\label{fig:method}
\end{center}
\vskip -0.05in
\end{figure}

\section{Methodology}
\subsection{Preliminary}
\label{sec:bg}
Given a large language model $M=\{\mathcal{E}_l,\mathcal{C}_l\}$ with a token set $T$ and hidden space $\mathbb{R}^m$, and an input sequence $t = [t_1, ..., t_K|t_k \in T]$, where $\mathcal{E}_l$ is the encoder, $\mathcal{C}_l$ is the logit projector, and $K$ represents the sequence length. The model generates the next token by:
\begin{equation}
\label{equ:gen}
t_{K+1} = M(t_{\le K}) = \mathcal{C}_l(\mathcal{E}_l(t_{\le K})),
\end{equation}
where $t_{K+1}$ is the next token and $h_{K} = \mathcal{E}_l(t_{\le K}) \in \mathbb{R}^m$ is the hidden state of the last token.

Indeed, LLMs generate tokens autoregressively, using the previous output token to predict the subsequent token.
This generation process continues until a stop condition is met, which may involve reaching a maximum token limit or generating a specific end-of-sequence token. 
Additionally, in modern LLMs, the Key-Value Cache (KV Cache) \citep{radford2018improving} technique is extensively utilized during inference to speed up attention map computations.

\vspace{-5pt}
\subsection{Jailbreak Detection}
\label{sec:det}
Most previous jailbreak detection methods either require proxy LLMs to analyze the model's output or involve multiple augmentations to the model's input, which are time-consuming and impractical for real-world applications.
Therefore, we propose training an efficient jailbreak detector that leverages the autoregressive generation properties of the model.
Specifically, as shown in the part 1 in \Cref{fig:method}, we train a gist token $t_g$ and a binary classifier $\mathcal{C}_d$, and obtain the predicted probability distribution $p_t$ of the text $t$ as follows:
\begin{equation}
\label{equ:det1}
\begin{aligned}
p_t &= \mathcal{C}_d(h_t) = \mathcal{C}_d(\mathcal{E}_l(t,t_g)),
\end{aligned}
\end{equation}
where $h_t$ represents the hidden state of the last token $t_g$. And then we obtain the detection results with $p_t$ as follows:
\begin{equation}
\label{equ:det2}
\mathop{\arg\max}\limits_c p_{t,c} =
\begin{cases} 0, \text{ not jailbroken,} \\ 1, \text{ jailbroken.}
\end{cases}
\end{equation}
We inject the $t_g$ token at the end of the sequence.
Since the keys and values of the previous tokens are cached during generation, the hidden state of $t_g$ can be computed efficiently based on the KV Cache.
For instance, for a sequence with a length of 2000, the cost of detecting jailbreak activities is approximately 1/1000 of the total generation time.
A simpler alternative would be to remove the gist token and directly use the hidden state of the last token to perform detection.
However, intuitively, the hidden state of the last token is used for generation and may not encapsulate the information relevant to the harmfulness of the response.
Therefore, we train a gist token designed to capture the harmfulness of the previous answer.
Additionally, we construct a dataset $\mathcal{D}_d = {(q_i,a_i,y_i)}_{i=1}^{|D_d|}$ to train our detector, where $q_i$ represents the question, $a_i$ represents the answer, and $y_i$ is the label indicating jailbreak activities.
We train the detector using naive cross-entropy loss, as follows:
\begin{equation}
\label{equ:detection}
\begin{aligned} t_g^*, \mathcal{C}_d^* = \mathop{\arg\min}\limits_{t_g,\mathcal{C}_d} \mathbb{E}_{(q_i,a_i,y_i) \sim \mathcal{D}_d} \left[ -\sum_{c=0}^1y_{i,c} \log\hat{p}_{i,c}\right],
\end{aligned}
\end{equation}
where $\hat{p}_i = \mathcal{C}_d(\mathcal{E}_l(q_i,a_i,t_g))$ represents the predicted jailbreak probability of jailbreak detector.

\subsection{Adaptive Defense Training}
\label{sec:sft}
Since detecting jailbreak activity is easier than directly defending against it, we build a test-time jailbreak defense system that transfers detection capability to defense capability that resembles the biological immune system.
When pathogens first enter the system, the body recognizes this invasion. In our approach, we treat jailbreak activities as pathogens and use the above detector to distinguish them from normal activities.
Once pathogens are identified, the organism will initiate an immune response and produce antibodies to neutralize the damage caused by antigens.
Following an immune response, the organism becomes immune to the specific antigen.
Similarly, when jailbreak activities are detected, our framework adds the detected jailbreak instructions along with a refusal response into jailbreak memory $\mathcal{M}_j$.
We then use $\mathcal{M}_j$ to supervise fine-tuning the model.
In this way, we progressively collect jailbreak data during the model testing process and enhance the defense capabilities of the model against various jailbreak attacks.
For normal instruction, our model does not alter its behavior but only incurs a slight time cost for detecting jailbreak activities.
Additionally, to prevent the model from becoming overly defensive against normal activities, we use the traditional question-answering (QA) dataset $\mathcal{D}_{qa}$, to regularize the model during training.

Furthermore, we adopt the concept of \textbf{test-time adaptation (TTA)} \citep{wang2021tent} to train our jailbreak detector with \Cref{equ:detection} while detecting jailbreak behaviors.
Specifically, we use detected jailbreak instructions along with their corresponding answers as jailbreak QA pairs, and jailbreak instructions with refusal responses as normal QA pairs.
We then append them to the detection memory, denoted as $\mathcal{M}_d$, and use $\mathcal{M}_d$ to train our detector by \Cref{equ:detection}.
Additionally, we also use the detection dataset, denoted as $\mathcal{D}_d$, for regularization training.

\subsection{Overall Framework}
\label{sec:dec}
Directly combining the above detection and defense training strategy comes with a drawback: the detector and defense training share a set of parameters (i.e., parameters in $\mathcal{E}_l$). The updates to model parameters by defense training are likely to impair the detector. To address this issue, we propose decoupling the detector and defense training.
For detection, we utilize the hidden state of the intermediate layer, rather than the last layer, to perform detection.
For defense training, we apply the LoRA module \citep{hu2022lora} to the layers behind the intermediate detection layer, treating them as trainable parameters, as shown in part 1 of \Cref{fig:method}.
We ensure that parameter updates to the detector and the defense training do not interfere with each other in this way.
After that, we obtain the overall pipeline of TIM. The details of our method can be found in \Cref{alg:TIM} for reference.

\section{Experiments}
\subsection{Setup}

\textbf{$\triangleright$ Jailbreak Attack/Defense Methods}. We evaluate our defense methods against various jailbreak attack methods. For experiments on MLLMs, we choose Figstep \citep{gong2023figstep} and MM-SafetyBench \citep{liu2024mm}.
For experiments on LLMs, we utilize I-FSJ and GCG (in the Appendix) as the jailbreak attack method.
For jailbreak defense methods, we consider FSD \citep{gong2023figstep}, Adashield \citep{wang2024adashield}, and VLGuard \citep{zong2024safety}.
Additionally, we introduce another baseline, TIM (w/o gist), which is identical to our method but uses the final hidden state of the last token for detection.
To assess the impact of our defense training on detection, we report results for TIM (w/o adapt.), where no defense training and optimization occur during testing.
Linear Probing (LP) represents a method that neither uses the gist token nor adapts during testing (i.e., LLMs with a linear probing binary detector on the last generated token). Furthermore, we compare our detector against detection baselines, including Self Defense \citep{phute2023llm} and LVLM-LP \citep{zhao2024first}, in LLM experiments.

\textbf{$\triangleright$ Metrics.}
We evaluate jailbreak methods from two perspectives: the effectiveness of defense against jailbreak attacks and the model's ability to respond to normal instructions. For evaluating the effectiveness of defense against jailbreak attacks, we adopt the Attack Success Rate (ASR) as a metric, as is common in most studies \citep{wang2024adashield,chao2023jailbreaking}. We define ASR as the proportion of jailbreak instructions that are not rejected, relative to all the jailbreak instructions. For the response set $R_j$ of the jailbreak dataset $\mathcal{D}_j$, ASR is calculated as follows:
\begin{equation}
    \begin{aligned}
        ASR = \frac{|R_j|-\sum_{r\in R_j} isReject(r)}{|R_j|}
        , \text{where}\ isReject(r)=
        \begin{cases}
        0, r\text{ is rejection,} \\
        1, r\text{ is not rejection.}
    \end{cases}
    \end{aligned}
\end{equation}
We employ prefix matching to determine whether a response is rejected. Specifically, we compile a set of rejection prefixes. If the model’s response matches any prefix in the rejection set, we consider the instruction rejected. The rejection prefixes employed are listed in \Cref{sec:eval}. Since our method aims to incrementally enhance the model’s security capabilities, we also report another metric, ASR-50, which calculates ASR for jailbreak samples in the last 50\% of the test sequences. This reflects the model’s performance after it has learned to defend against jailbreak attacks.
Although defense methods improve the model’s ability to reject malicious instructions, they may also cause the model to reject an excessive number of normal queries. Thus, we use the Over-Defense Rate (ODR) to assess the model’s ability to respond to clean instructions. For the response set $R_n$ of the normal dataset $\mathcal{D}_n$, ODR is calculated as follows:
\begin{equation}
    \begin{aligned}
        ODR = \frac{\sum_{r\in R_n}isReject(r)}{|R_n|}.
    \end{aligned}
\end{equation}
Additionally, to evaluate the detector's performance, we report the Accuracy (ACC), True Positive Rate (TPR), and False Positive Rate (FPR) \citep{swets1988measuring}.
Moreover, we provide the details of our detection dataset, experiment setup and the introduction of our baselines in the \Cref{sec:details}.

\subsection{Main Results}
\begin{table}[h]
\centering
\caption{The experimental results under Figstep \citep{gong2023figstep}. TIM's ASR is reported in the format of ASR/ASR-50 (same in the subsequent manuscript).}
\label{tab:figstep}
\resizebox{0.75\linewidth}{!}{
\begin{tabular}{lcc|cc|cc}
\toprule
\multirow{2}{*}{Methods} & \multicolumn{2}{c|}{LLaVA-v1.6-Vicuna-7B} & \multicolumn{2}{c}{LLaVA-v1.6-Mistral-7B} & \multicolumn{2}{c}{LLaVA-v1.6-Vicuna-13B}  \\
                  &    ASR ($\downarrow$)       &  ODR ($\downarrow$)       &    ASR ($\downarrow$)      &   ODR ($\downarrow$)  &    ASR ($\downarrow$)      &   ODR ($\downarrow$) \\ \hline
               Vanilla   &   100.0        & 0.0         & 100.0          & 0.0  &100.0 &0.0           \\ 
               FSD \citep{gong2023figstep}   &   100.0        &  0.0        &   100.0        &   0.0  &100.0 & 0.0     \\
               Adashield \citep{wang2024adashield}  &   0.0        &   14.0       &    0.0   &  7.2 & 0.0 &  51.2      \\
               VLGuard \citep{zong2024safety}   &  0.0         & 7.0         &   0.0        &    1.8  & 0.0 & 5.2    \\
               TIM (w/o gist)  &   1.6         &  0.0        &     0.4      &  0.4  &0.8 &1.6       \\
               TIM   &   1.4/0.0        &   0.0       &    0.6/0.0       &   0.0 & 1.8/0.0 & 0.4  \\  \bottomrule  
\end{tabular}
}
\vskip -0.15in
\end{table}

\begin{table}[htbp]
\caption{The experimental results under the MM-SafetyBench \citep{liu2024mm}.}
\label{tab:MM}
\centering
\resizebox{0.55\textwidth}{!}{
\begin{tabular}{lcc|cc} \toprule
\multirow{2}{*}{Methods} & \multicolumn{2}{c}{LLaVA-v1.6-Vicuna-7B} & \multicolumn{2}{c}{LLaVA-v1.6-Vicuna-13B} \\ 
 &     ASR ($\downarrow$)      &    ODR ($\downarrow$)      &   ASR ($\downarrow$)        &     ODR ($\downarrow$)     \\ \midrule
Vanilla &    99.8       &   0.2       &     100.0      &     0.4     \\
FSD \citep{gong2023figstep} &      99.8     &    0.2      &   99.7        &      0.0    \\
Adashield \citep{wang2024adashield} &     7.0      &  14.0        &    43.8       &  51.5        \\
VLGuard \citep{zong2024safety} &       1.4    &   6.5       &   0.2        &  4.7        \\
TIM (w/o gist) &      1.4     &   10.7       &   3.0        &  3.8        \\
TIM &     1.0/0.0      &    2.3      &    4.8/0.0       &     0.4    \\ \bottomrule
\end{tabular}
}
\vskip -0.1in
\end{table}

\textbf{$\triangleright$ Jailbreak Defense.}
To evaluate the effectiveness of our method, we report the results on Figstep and MM-SafetyBench in \Cref{tab:MM,tab:figstep}. As shown in the tables, Adashield demonstrates strong defensive capabilities, especially against Figstep, where it reduces the ASR to 0\%.
Similarly, the ASR on MM-SafetyBench is reduced to 7\% by Adashield.
Despite its effectiveness, Adashield suffers from a noticeable over-defense phenomenon with normal samples, with over 5\% of them being rejected.
After training on a specially designed dataset, VLGuard shows relatively excellent performance, achieving almost 0\% ASR against jailbreak samples but still show over-rejects to normal samples.
Compared to VLGuard, our method can gradually learn to reject jailbreak attacks during testing without any prior targeted training.
It achieves an ASR of less than 2\% at most experiments, and, among all the effective jailbreak attack defense methods, our approach causes the least damage to the model's ability to respond to normal queries (i.e., ODR from 0.2\% to 2.3\% on MM-SafetyBench with LLaVA-v1.6-Vicuna-7B as backbone and nearly 0\% on others).
From the ASR, we can draw a conclusion that our method only requires a few jailbreak samples to learn how to reject such types of jailbreak attacks (on the Figstep dataset, this number is less than 10).
Since our method progressively enhances the model's defensive capabilities during testing, we believe that the ASR-50 metric better reflects the true effectiveness of our approach. Our method achieved 0\% ASR-50 across all jailbreak attack datasets, indicating that, with continuous optimization, our model can achieve complete defense against individual attacks.
Moreover, \Cref{tab:ifsj} shows the results for the text-based attack. Our method is also effective at defending against I-FSJ, a jailbreak method that only uses the language modality. TIM not only achieves an ASR-50 of 0\% but also reduces the model's ODR.
\begin{table}[h]
\caption{The experimental results under text-based attack, I-FSJ \citep{zheng2024improved}.}
\label{tab:ifsj}
\centering
\resizebox{0.7\textwidth}{!}{
\begin{tabular}{l|ccc|ccc} \toprule
\multirow{2}{*}{Methods} & \multicolumn{3}{c|}{LLaMA2-7B-chat} & \multicolumn{3}{c}{LLaMA3-8B-Instruct} \\ 
 &     ASR ($\downarrow$)      &    ODR ($\downarrow$) & TPR ($\uparrow$)      &   ASR ($\downarrow$)        &     ODR ($\downarrow$)  & TPR ($\uparrow$)   \\ \midrule
Vanilla &   99.2        &   5.5   & -    &   94.3        &   0.2  & -     \\
Retokenization (20\%) & 97.5 & 8.3& - &83.0 & 0.2& - \\
SmoothLLM (insert 20\%) & 76.6& 26.7&- &100.0 &0.4 &- \\
SmoothLLM (swap 20\%) & 93.4& 55.8 & -& 60.0 & 1.8 &- \\
SmoothLLM (patch 20\%) &80.9 & 27.5 &- & 57.4& 6.4 &- \\
TIM (w/o adapt.) & - & - & 98.9 & - &- &18.2 \\
TIM (w/o gist) &      0.6     &    4.9  & 100.0    & 12.7          &  19.7  & 1.5      \\
TIM &    2.6/0.0       &  0.6   &100.0     &  1.0/0.0         &   0.2 & 40.0   \\ \bottomrule   
\end{tabular}
}
\end{table}

\begin{figure}[htbp]
    \centering
    \vskip -0.1in
    \includegraphics[width=\linewidth]{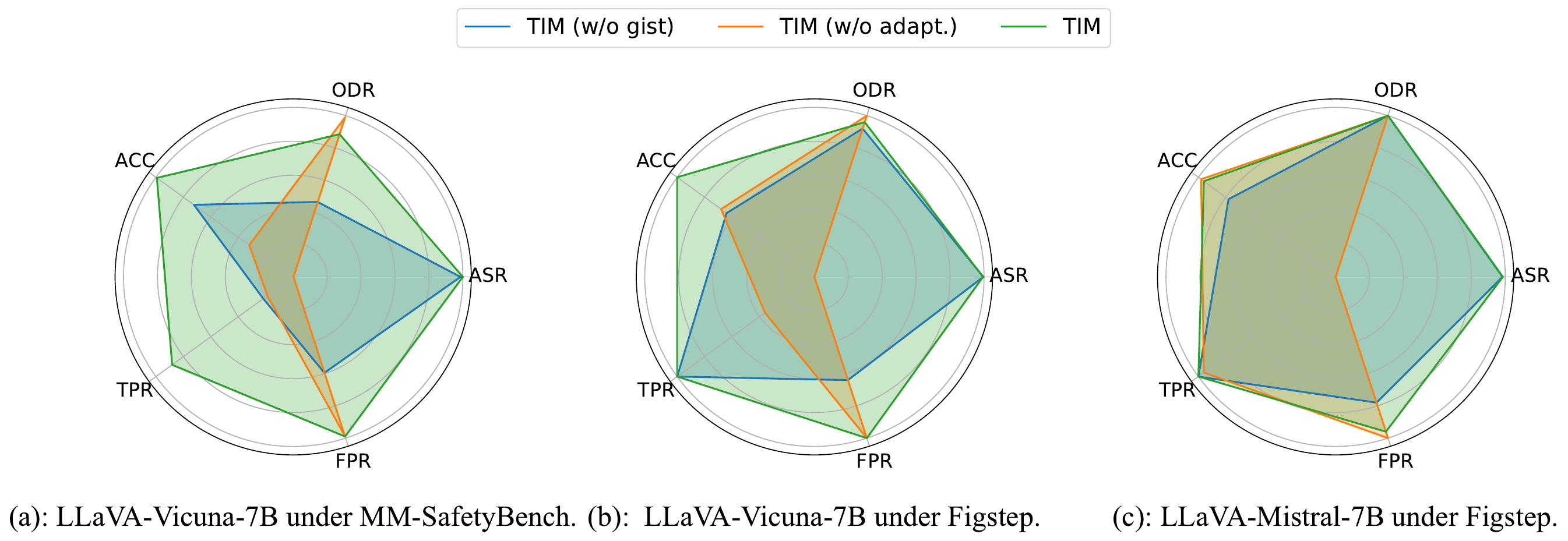}
    \caption{Performance of different variants of the proposed method. All metrics are normalized, and the methods with larger areas have better performance.}
    \label{fig:ablation}
    \vskip -0.1in
\end{figure}

\textbf{$\triangleright$ Jailbreak Detection.}
Next, we analyze the role of our jailbreak detector from two perspectives: 1) What advantages does our detector's design offer compared to TIM (w/o gist)? 2) How does training the detector during testing enhance the effectiveness of our framework?
First, addressing the initial question, the results in \Cref{tab:detection} show that TIM (w/o adapt.) exhibits clear improvements over LP in three metrics: Accuracy, TPR, and FPR. This improvement is primarily attributed to our introduction of the gist token, which is specifically designed to extract malicious information from previously generated sequences, rather than relying solely on the output of the last token for classification. This strategy has improved the expressive capacity of our detector.

\begin{wrapfigure}{r}{0.5\textwidth}
\vspace{-0.5cm}
\begin{minipage}{\linewidth}
\centering
\captionof{table}{The detection performance under I-FSJ attack.}
\label{tab:detection}
\resizebox{\linewidth}{!}{
\begin{tabular}{lcc|c}
\toprule
Methods   & ACC ($\uparrow$) & TPR ($\uparrow$) & FPR ($\downarrow$) \\ \midrule
Self Defense \citep{phute2023llm}  & 64.4  & 42.9  & 14.2 \\
LVLM-LP \citep{zhao2024first}       & 67.7  & 36.3  & 0.8  \\
LP     & 88.5  & 77.4  & 0.7  \\
TIM (w/o adapt.)
& 99.1  & 98.9  & 0.6  \\
TIM (w/o gist) & 99.4  & 100.0 & 0.6  \\
TIM           & 99.9  & 100.0 & 0.1  \\ \bottomrule
\end{tabular}
}
\end{minipage}
\vspace{-0.3cm}
\end{wrapfigure}

Secondly, the performance of the detector is shown in \Cref{fig:ablation}. It is evident that TIM (w/o gist) exhibits a significant increase in FPR compared to TIM, suggesting that it misclassifies more normal samples as jailbreak samples. One consequence of this issue is the use of more normal samples in defense training, which leads to an increase in the model's ODR, as shown in the results in \Cref{tab:MM,tab:ifsj}.
The cause of this issue arises from the detector sharing parameters with the defense training. The parameters' update during defense training will affect the performance of the detector. TIM resolves this issue by decoupling the defense training from the jailbreak detector by separating parameters.

\subsection{Additional Analysis}
\begin{wrapfigure}{r}{0.5\columnwidth}
\centering
\includegraphics[width=\linewidth]{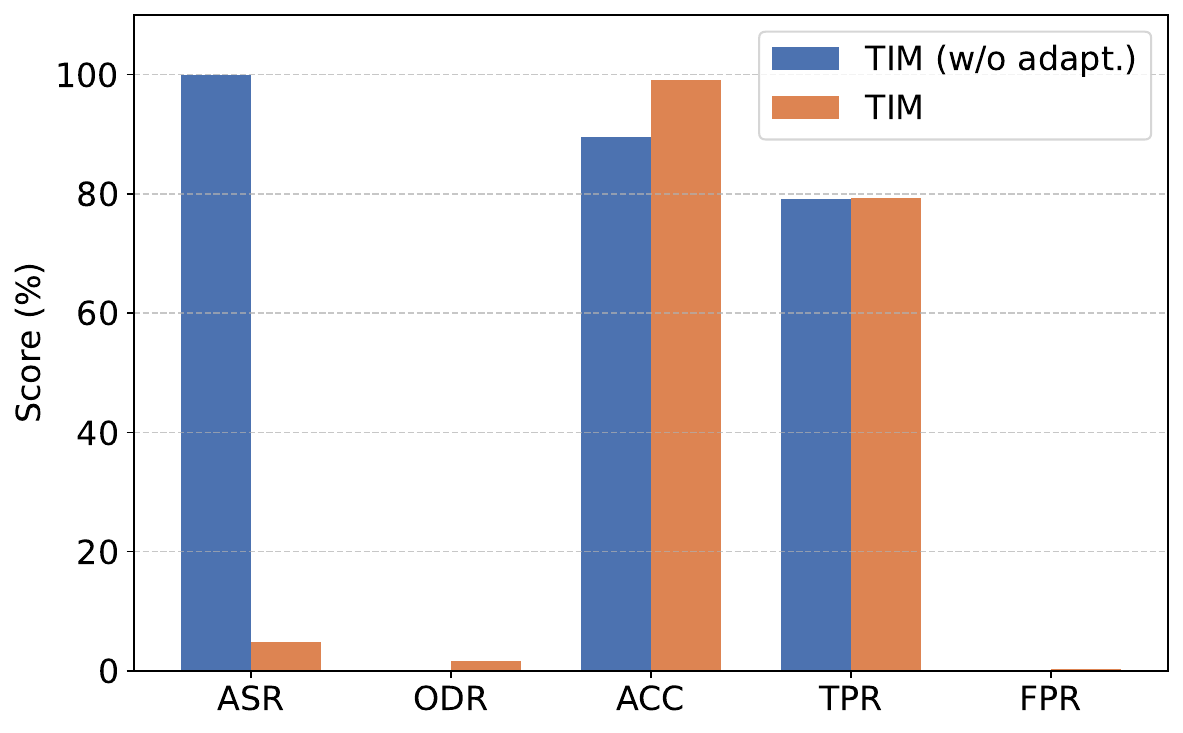}
\caption{Results under hybrid jailbreak attack. We randomly selected 300 jailbreak samples from MM-SafetyBench and 300 from Figstep, combining them into a new jailbreak dataset.}
\label{fig:mixed}
\vspace{-0.15in}
\end{wrapfigure}

In real-world scenarios, the situations encountered by models can be both complex and diverse.
Therefore, we conduct additional experiments to directly assess the robustness of our method in complex scenarios.
\textit{The results of transferability, continually changing jailbreak, and GCG attack are provided in the \Cref{sec:add_res}.}

\textbf{$\triangleright$ Sensitivity to the Detector.}
The ability of our method to resist jailbreak attacks intuitively depends on the detector's effectiveness at identifying such attacks.
As shown in \Cref{tab:ifsj}, our detector exhibited a relatively lower TPR under certain extreme conditions. Specifically, TIM (w/o adapt.) detected only 18.2\% of jailbreak activities; however, with test-time adaptation of the detector, TIM significantly improved detection performance, achieving a TPR of 40\%.
We hypothesize that this reduced detection efficacy occurs because I-FSJ requires eight context examples to successfully jailbreak LLaMA3-8B-Instruct, resulting in a substantial discrepancy between the token lengths encountered during detector training and those in testing scenarios. 
The average token lengths for instructions and answers during detector training are 13 and 271, respectively, whereas the average token length for jailbreak instructions using I-FSJ reaches 3061.
Despite this limitation, our method effectively resists attacks on LLaMA3, demonstrating robustness even when the detector's performance degrades.

\textbf{$\triangleright$ Results under Hybrid Jailbreak Attack.}
In deployment scenarios, attackers may employ multiple methods simultaneously to launch jailbreak attacks against the model. Accordingly, we designed experiments involving hybrid jailbreak attacks. The results, presented in \Cref{fig:mixed}, indicate that under our method, the ASR can still be reduced to a very low level, while the model's ability to respond to normal queries remains largely unaffected.

\begin{figure}[tbp]
    \centering
    \subfigure[The defense capabilities of our method.]{
        \includegraphics[width=0.4\textwidth]{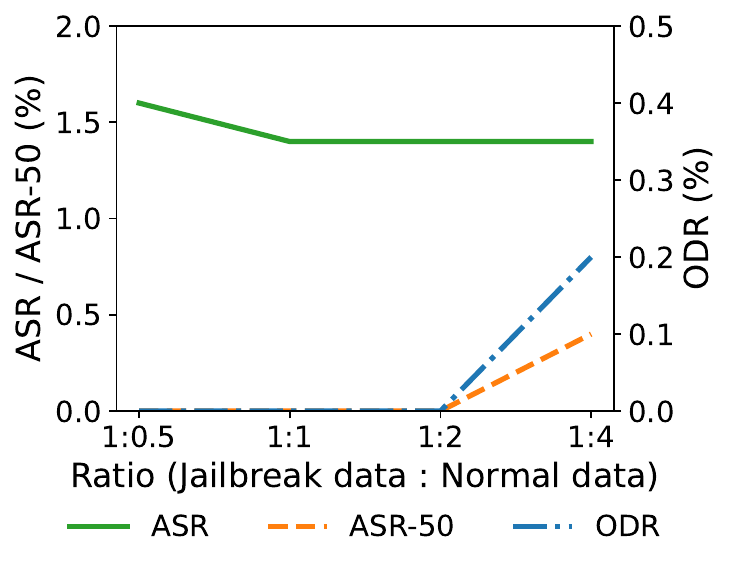}
    }
    \hspace{0.5cm} 
    \subfigure[The detection performance of our method.]{
        \includegraphics[width=0.4\textwidth]{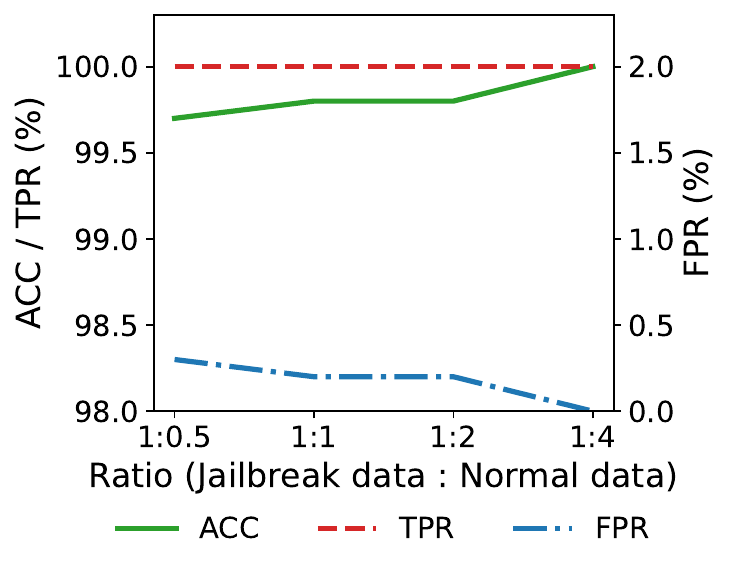}
    }
    \caption{Experimental results under different jailbreak data ratios.}
    \label{fig:ratio}
\end{figure}
\textbf{$\triangleright$ Results under Different Jailbreak Data Ratios.}
In practical applications, the proportion of jailbreak data within the model's test data is typically not fixed. The model may simultaneously receive a large number of jailbreak attack requests, or it might not encounter any jailbreak instructions for extended periods. Thus, we report the results of our method under varying proportions of jailbreak attack data in \Cref{fig:ratio}.
The results presented in the table demonstrate that our method achieves stable and effective performance across various proportions, both in terms of defending against jailbreak attacks and the detection performance of our detector.

\begin{table}[h]
\caption{Average inference cost (seconds) for each instruction. All experiments are conducted with I-FSJ jailbreak. The test samples are mixed with 520 normal samples and 520 jailbreak samples.}
\label{tab:cost}
\centering
\resizebox{0.7\textwidth}{!}{
\begin{tabular}{c|cc|cc}
\toprule
Vanilla&\multicolumn{2}{c|}{Detection}  & \multicolumn{2}{c}{Test-time Defense} \\ 
               LLaMA2-7B    & + TIM's Detector   &           + Self Defense        &   TIM    &  Training Inside  \\ \midrule
    7.18        &   7.21 (+0.4\%)       &  36.13   & 5.49      &   0.67 (12.2\%)    \\ \bottomrule
\end{tabular}

}
\end{table}

\textbf{$\triangleright$ Computation Cost Analysis}.
The computational cost of our method is reported in \Cref{tab:cost}.
As shown, our detector introduces a negligible overhead—\textit{only 0.4\% of the standard inference cost}—making it substantially more efficient than Self Defense \citep{phute2023llm}, which adopts a proxy LLM to analyze the generated output.
In addition, the training cost constitutes merely 12.2\% of the overall computational budget.
Overall, the inference time of TIM is lower than that of the vanilla model.
This is primarily because TIM generates short rejection responses to jailbreak attempts, rather than generating long malicious outputs.

\section{Conclusion}
In this paper, we address the challenge of defending against diverse jailbreak attacks.
We propose a universal test-time defense framework designed to dynamically detect jailbreak attacks during testing and utilize detected jailbreak instructions to defensively train the model.
To enhance jailbreak attack detection, we introduce a specialized gist token designed to extract harmful information from model responses with almost no additional cost, which is then classified using a binary classifier.
Furthermore, to minimize the impact of model updates on the detector, we decouple the detector from defense training, ensuring they operate on separate parameters and do not interfere with each other.
Extensive experiments demonstrate the efficacy of our method across a variety of scenarios.

\bibliography{main}
\bibliographystyle{plainnat}
\appendix

\section{The Details of Experimental Setup}
\label{sec:details}
\subsection{Dataset Construction}
\label{sec:dataset}
To construct the detection dataset, we initially collected original malicious instructions from AdvBench \citep{zou2023universal} and MM-SafetyBench \citep{liu2024mm}.
To obtain malicious answers, we employed Wizard-Vicuna-7B-Uncensored \citep{xu2024wizardlm}, a model without safety alignment, to generate answers.
To obtain refusal answers, we utilized LLaMA2-13B-chat to generate answers with various refusal prefixes.
We employed GPT4-LLM-Cleaned \citep{peng2023instruction} and LLaVA-Instruct-150K \citep{liu2023llava} as clean instructions for LLMs and MLLMs, respectively.
Furthermore, to generate clean answers, we utilized LLaMA2-7B-chat and LLaVA-v1.6-Vicuna-7B for GPT4-LLM-Cleaned and LLaVA-Instruct-150K, respectively.
Our detection dataset comprises four parts: 1) malicious instructions with malicious answers, classified as jailbroken; 2) malicious instructions with refusal answers, classified as not jailbroken; 3) clean instructions with clean answers, classified as not jailbroken; 4) clean instructions with malicious answers, classified as jailbroken.
The primary focus of the dataset is to determine whether the answer is harmful, rather than assessing the harm of the instruction itself.
For the visual question-answering (VQA) dataset, since the original malicious instructions lack images, we randomly selected images from the COCO dataset \citep{lin2014microsoft} for them.
It is important to note that our malicious instructions are original and unaffected by jailbreak attacks, meaning we do not use jailbreak-processed instructions during detector training.
For the evaluation dataset, we combine normal QA/VQA instructions from GPT4-LLM-Cleaned/LLaVA-Instruct-150K with jailbreak instructions to simulate real deployment environments in experiments on LLMs/MLLMs.

\subsection{Baselines}
\textbf{Figstep} \citep{gong2023figstep} conceals harmful content within text prompts using typography, embedding it into blank images to circumvent text-modality safety alignments.

\textbf{MM-SafetyBench} \citep{liu2024mm} initially generates a malicious background image using harmful keywords from jailbreak prompts and subsequently converts text-based harmful content into images using topography.

\textbf{I-FSJ} \citep{zheng2024improved}, based on in-context jailbreak \citep{wei2023jailbreak}, aims to induce the model to generate harmful content through several jailbreak demonstrations.
Additionally, I-FSJ employs system tokens to enhance its attack capabilities.
Furthermore, a greedy search is used to select the optimal demonstration from the datasets.

\textbf{GCG} \citep{zou2023universal} is a white-box method utilizing an adversarial text suffix to jailbreak LLMs. 

\textbf{FSD} \citep{gong2023figstep} is a defense method that introduces a specific system prompt, reminding the model to focus on malicious text within images.

\textbf{Adashield} \citep{wang2024adashield} is a test-time alignment method proposing the addition of a defense prompt following the input text prompt. The defense prompts can be static or adaptive, which are called Adashield-S or Adashield-A, respectively.
We consider Adashield-S in our experiments.

\textbf{VLGuard} \citep{zong2024safety} is a training-time alignment method that involves additional safety fine-tuning on a specific dataset.
It constructs a safety instruction tuning dataset containing malicious images to defend against structure-based jailbreak methods like Figstep and MM-SafetyBench.
Unlike VLGuard, our detector's training dataset contains no prior knowledge of the jailbreak attack method (e.g., malicious images).

\subsection{Experimental Details}
For MLLM experiments, we select LLaVA-v1.6-Vicuna-7B/13B \citep{chiang2023vicuna} and LLaVA-v1.6-Mistral-7B \citep{liu2023llava,liu2024llavanext,liu2023improved,jiang2023mistral} as the base models.
For LLM experiments, we use LLaMA2-7B-chat and LLaMA3-8B-Instruct \citep{touvron2023llama} as the base model.
The weights for all base models are sourced from Hugging Face.
We set the learning rate, number of epochs, and batch size for detector training to 1e-3, 5, and 32, respectively.
We use the Adam optimizer \citep{kingma2014adam} for defense training, setting the learning rates to 0.001 for MLLMs and 0.002 for LLMs.
We apply LoRA \citep{hu2022lora} with a rank of 16 to the query and value matrix in the last 15 transformer blocks.
The regularization batch size is set to 40, while the batch sizes for refusal training and detector training during test time are set to 1 and 6, respectively.
Furthermore, during jailbreak activity detection, we train the defense capabilities and the detector for 1 and 5 steps, respectively.
We incorporate an equal mix of jailbreak instructions and clean instructions in the test data.
All the experiments are conducted on 4*RTX A6000.

\subsection{The jailbreak Evaluation}
\label{sec:eval}
For the refusal judgment, we use the prefixes match method.
We think the model refused to answer if the respones start with the following prefixes.

\begin{minipage}[]{0.4\linewidth}
{\fcolorbox{black}{gray!10}{\parbox{\linewidth}{I'm sorry \\
    Sorry \\
    I am sorry \\
    I apologize\\
    As an \\
    As a responsible\\
    I'm an \\
    I'm just \\
    As an Assistant \\
    If you have any other non-malicious requests
    }
}} 
\end{minipage}
\hspace{5mm}
\begin{minipage}[]{0.4\linewidth}
\centering
{\fcolorbox{black}{gray!10}{\parbox{\linewidth}{
    I do not \\
    I cannot \\
    As a language model \\
    I'm really sorry\\
    My apologies\\
    I'm not able to provide \\
    I am really sorry\\
    I can't provide\\
    I can't assist \\
    
    }
}} 
\end{minipage}

\section{Additional Results}
\label{sec:add_res}
\noindent 
\begin{wraptable}{r}{0.4\textwidth} 
    \centering
    \caption{ASR(\%) under continual changing environments.}
    \label{tab:continual}
    \begin{tabular}{ccc}
    \toprule
    \multicolumn{3}{c}{Attack Order ($\longrightarrow$)} \\
    Figstep & MM-SafetyBench & Figstep \\ 
    \midrule
    1.4 & 6.6 & 0.0 \\
    \bottomrule
    \end{tabular}
\end{wraptable}

\textbf{Robustness against continual changing attack}. To validate the effectiveness of our method under continuous exposure to various forms of jailbreak attacks, we conducted experiments as shown in \Cref{tab:continual}. We selected 500 different samples for each type of jailbreak attack and conducted the attacks in varying orders. As can be seen, even after undergoing the MM-SafetyBench attack, our method still maintains good defensive performance during the second exposure to the Figstep attack, without experiencing catastrophic forgetting.

\begin{table*}[ht]
\caption{The transferability results. We first adopt TIM on the source jailbreak attack. Then, we freeze the fine-tuned model and evaluate it on the target attack. We report the ASR while adopting the LLaVA-v1.6-Vicuna-7B as the backbone. The numbers in brackets represent the changes of ASR compared to the Vanilla Model.}
\label{tab:transfer}
\centering
\begin{tabular}{cc} \toprule
Figstep $\longrightarrow$ MM-SafetyBench& MM-SafetyBench $\longrightarrow$ Figstep \\ \midrule
84.3 (-15.5) & 0.0 (-100.0) \\ \bottomrule
\end{tabular}
\end{table*}

\textbf{Transferability of defense training}. We demonstrate the static transferability of the fine-tuned model in \Cref{tab:transfer}. It is effective when migrating from a more complex attack (MM) to a simpler one (Figstep), but its effectiveness is limited in the reverse direction. However, it's worth noting that our method is an online adaptive defense method. New types of jailbreaks will be adaptively defended against as they emerge.

\begin{table*}[ht]
\caption{Experimental Results under GCG jailbreak attacks.}
\label{tab:gcg}
\centering
\resizebox{0.4\textwidth}{!}{
\begin{tabular}{l|cc}
\toprule
 & ASR & ODR \\ \midrule
LLaMA2-7B-chat & 21.5 & 0.2 \\
+TIM & 7.7 (-13.8\%) & 2.7 (+2.5\%)\\ \bottomrule
\end{tabular}
}
\end{table*}

\begin{figure*}[h]
\centering
\begin{minipage}[b]{.35\linewidth}
    \centering
    \subfigure[Accumulated ASR]{\includegraphics[width=\linewidth]{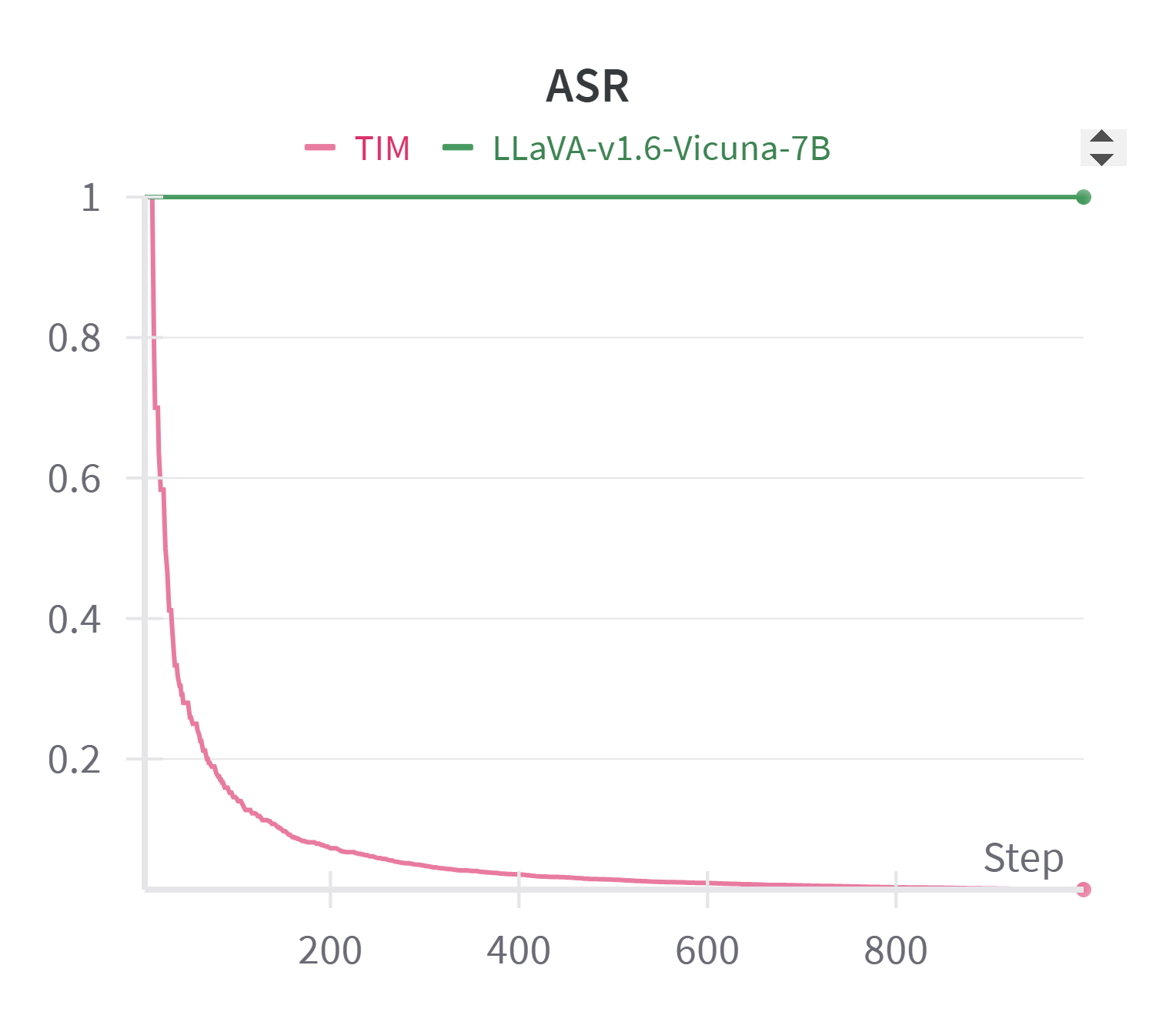}}\\
    \subfigure[Accumulated TPR]{\includegraphics[width=\linewidth]{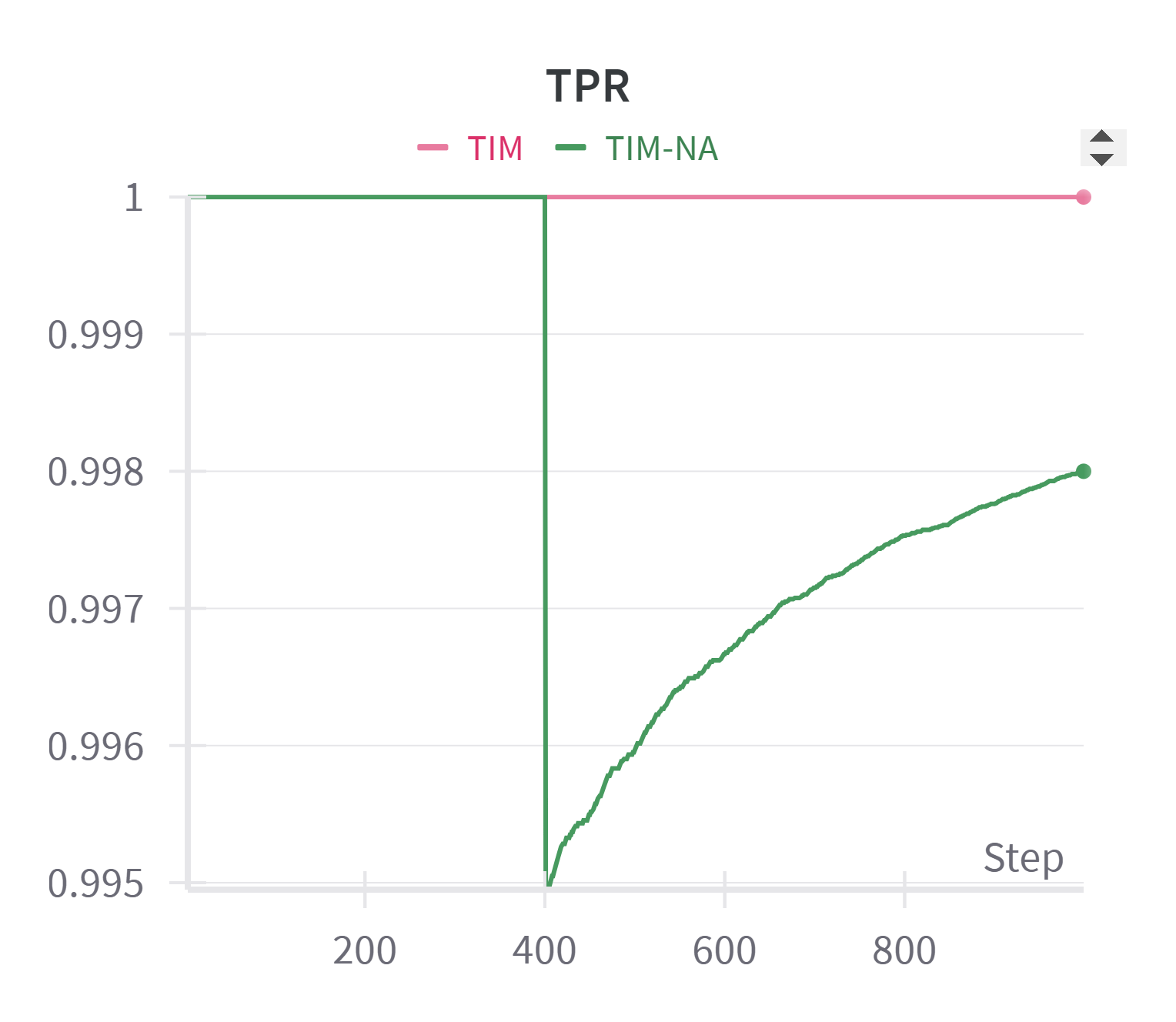}}
\end{minipage}
\begin{minipage}[b]{.35\linewidth}
    \centering
    \subfigure[Accumulated ODR]{\includegraphics[width=\linewidth]{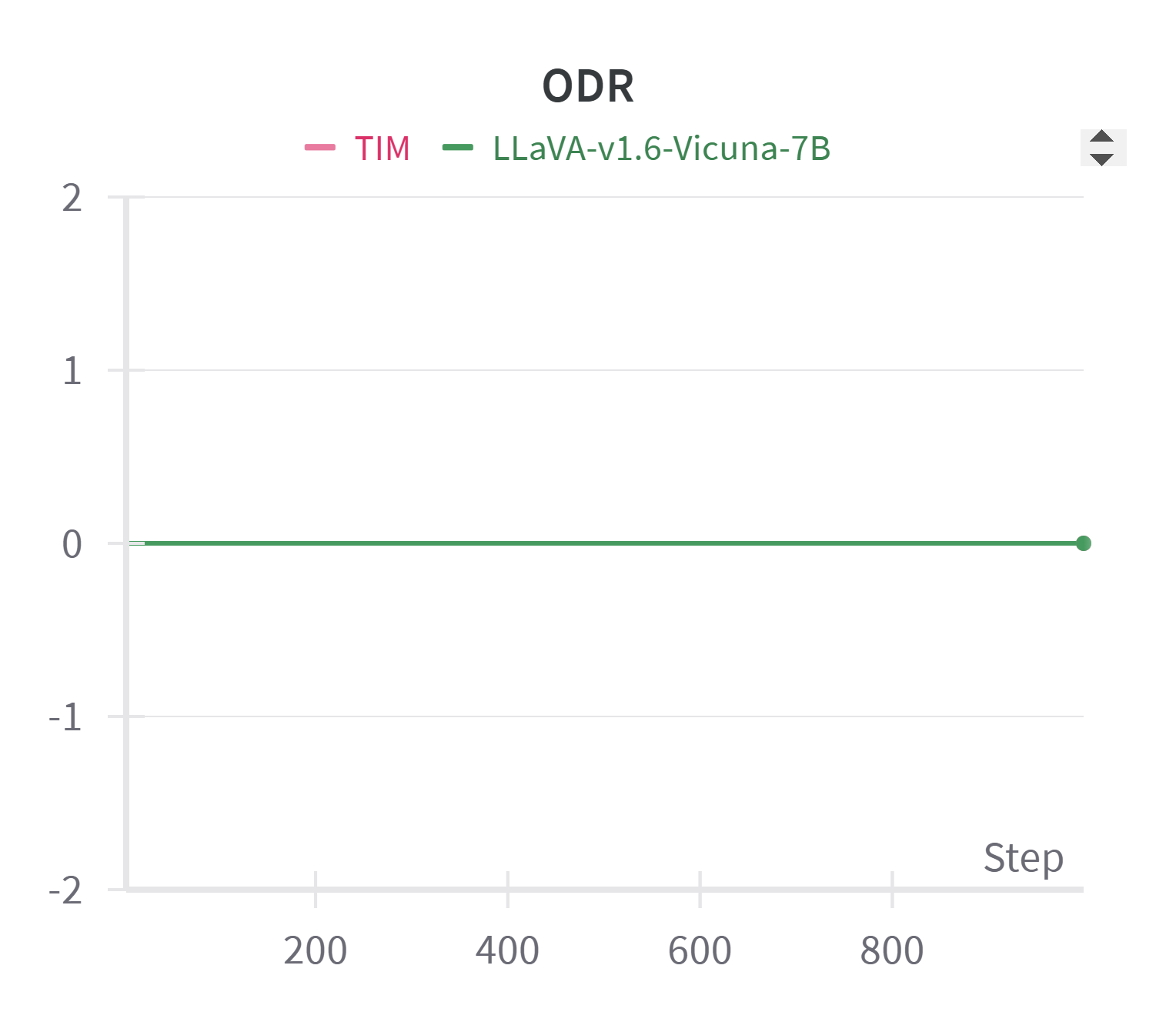}}\\
    \subfigure[Accumulated FPR]{\includegraphics[width=\linewidth]{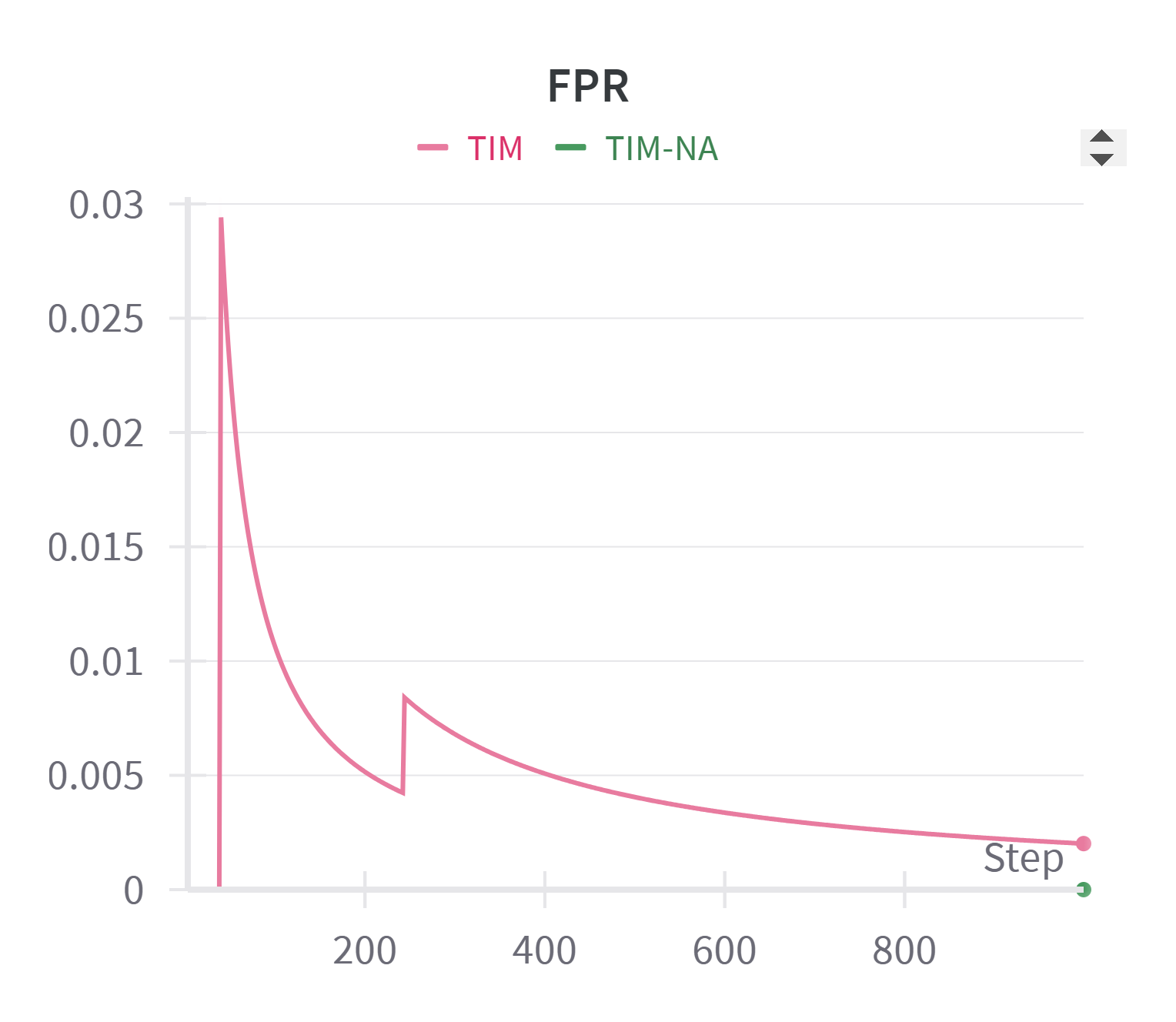}}
\end{minipage}
\caption{Changes in metrics during the test process against Figstep. TIM-NA represents TIM (w/o adapt.)}
\label{fig:figstep_test}
\end{figure*}

\begin{figure*}[h]
\centering
\begin{minipage}[b]{.35\linewidth}
    \centering
    \subfigure[Accumulated ASR]{\includegraphics[width=\linewidth]{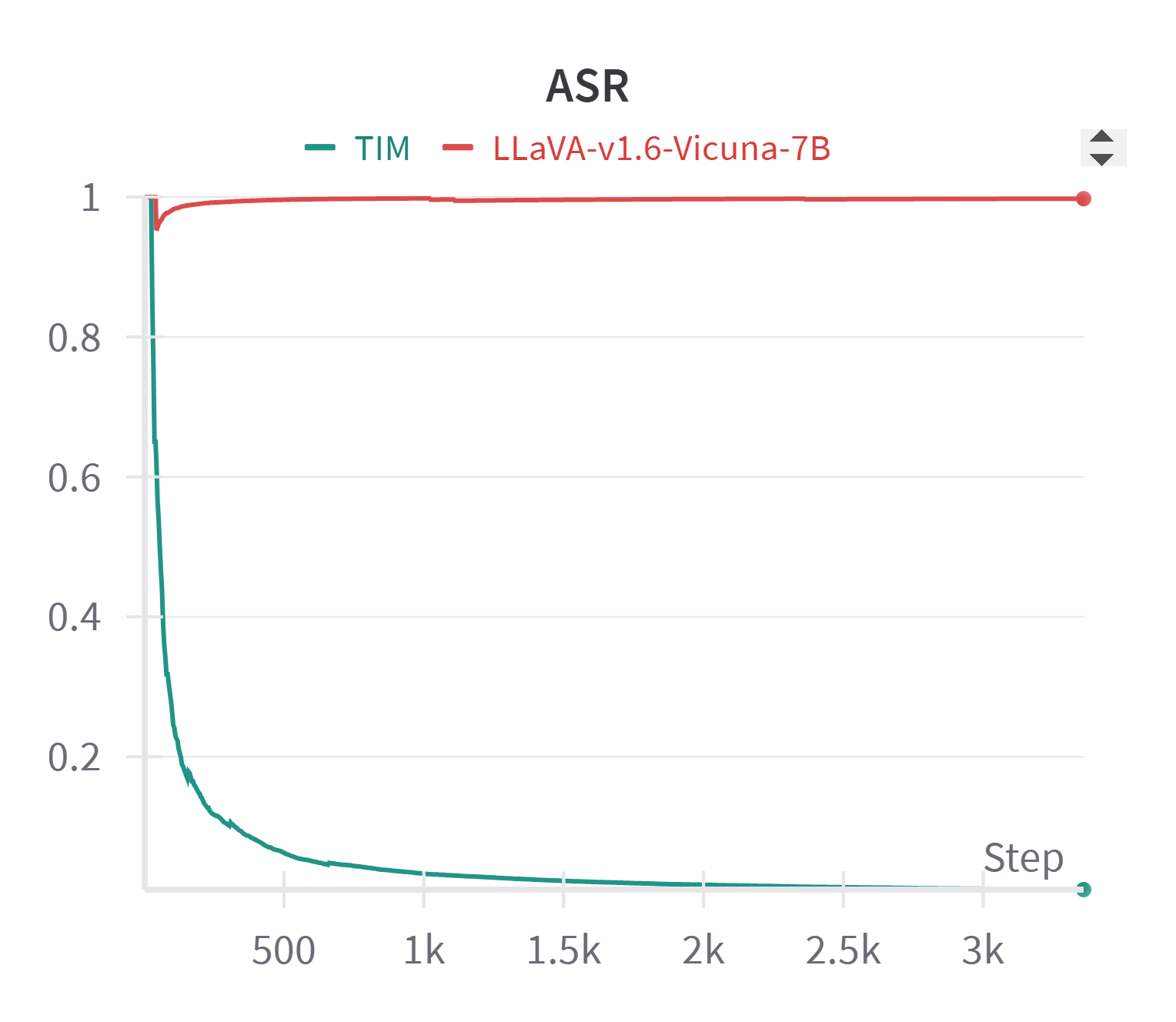}}\\
    \subfigure[Accumulated TPR]{\includegraphics[width=\linewidth]{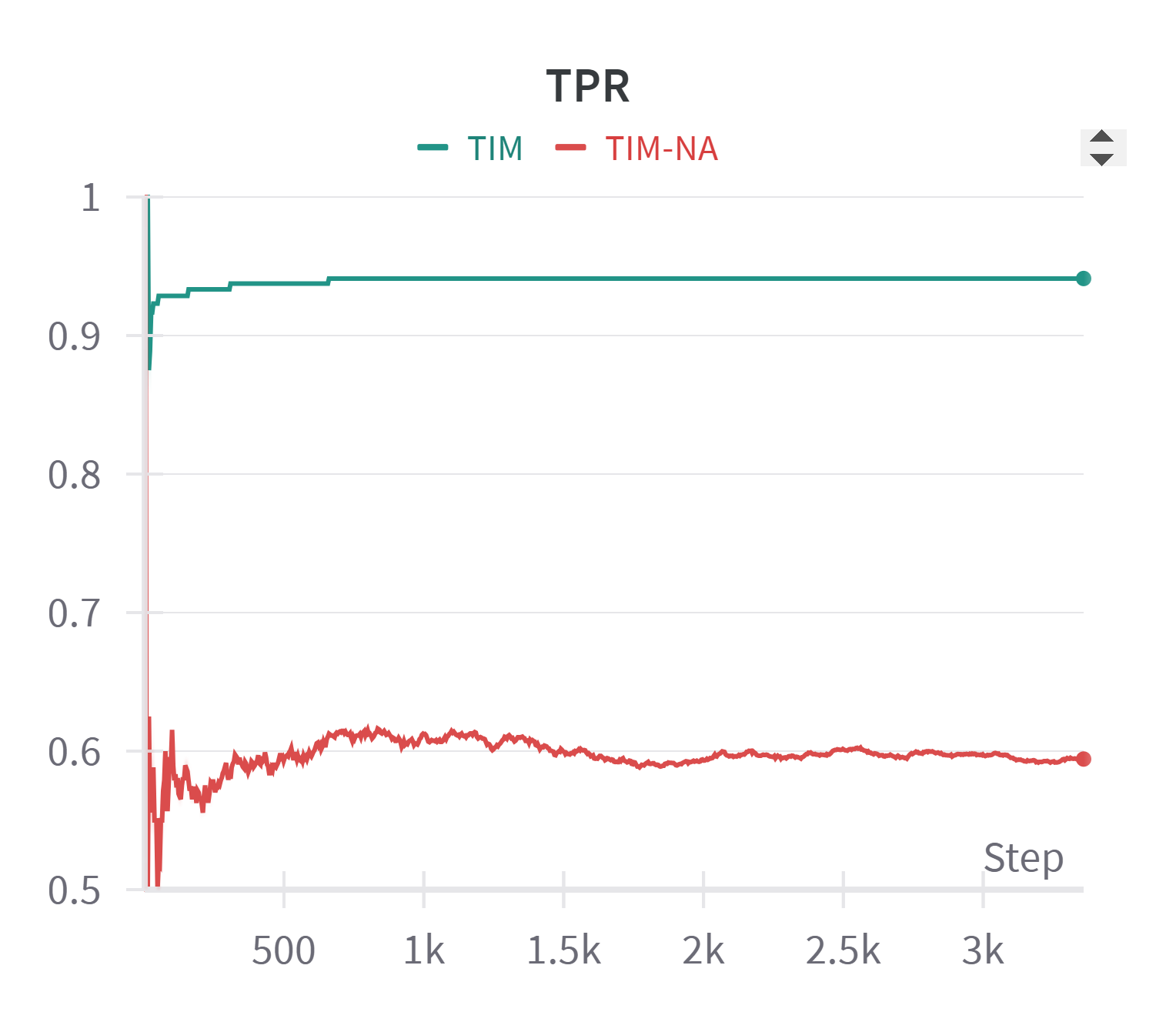}}
\end{minipage}
\begin{minipage}[b]{.35\linewidth}
    \centering
    \subfigure[Accumulated ODR]{\includegraphics[width=\linewidth]{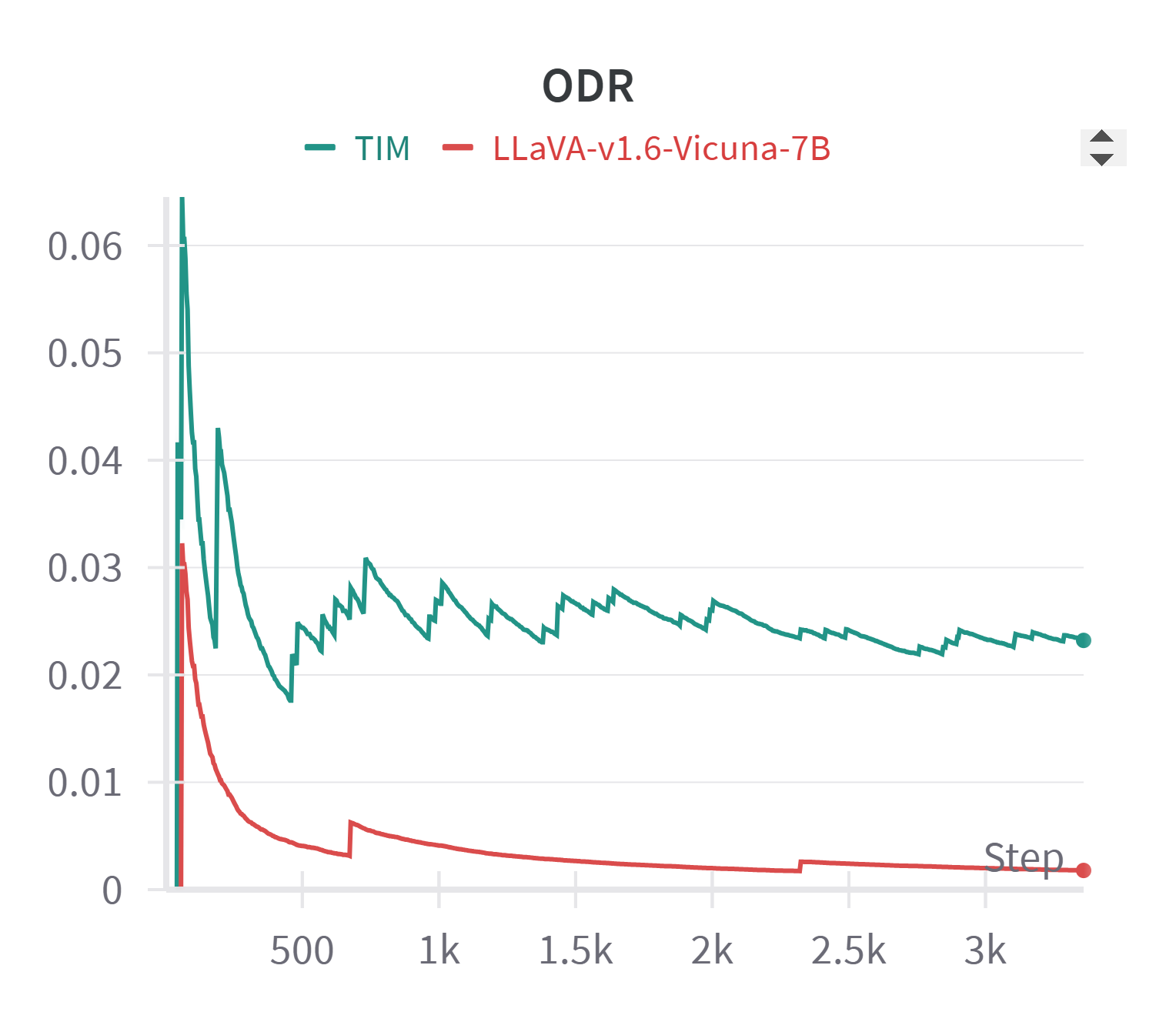}}\\
    \subfigure[Accumulated FPR]{\includegraphics[width=\linewidth]{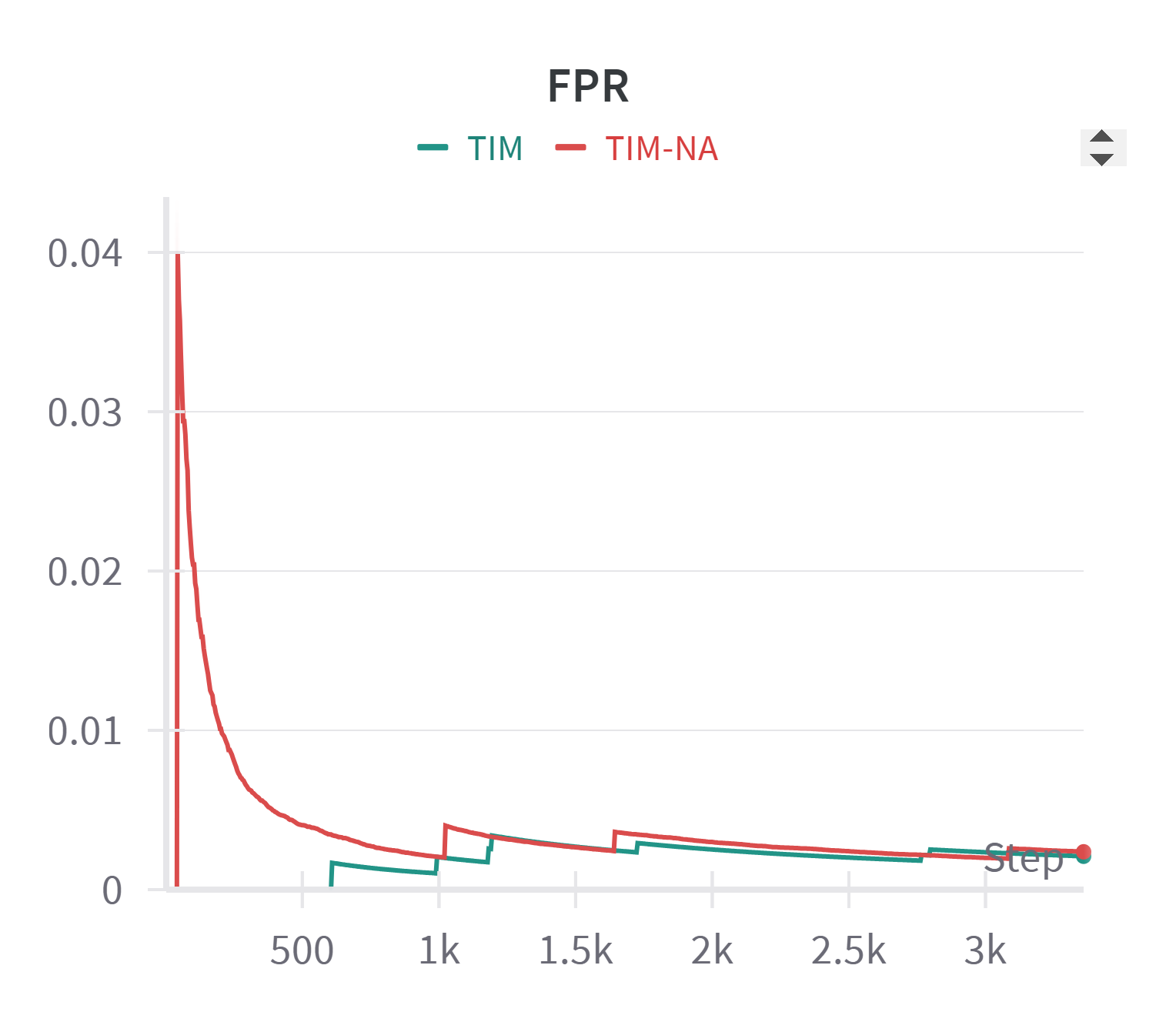}}
\end{minipage}
\caption{Changes in metrics during the testing against MM-SafetyBench. TIM-NA represents TIM (w/o adapt.)}
\label{fig:mm_test}
\end{figure*}

\textbf{Results under GCG attack}. We supplemented the results of the white-box attack, GCG, in \Cref{tab:gcg}. TIM decreased the ASR from 21.5\% to 7.7\%, demonstrating its effectiveness against GCG.

\textbf{Performance curve during testing}. To demonstrate the performance of our method as the test progresses, we report the relevant indicators in the \Cref{fig:figstep_test,fig:mm_test}. As can be seen, as the test progresses, the ASR of our method continues to decrease, indicating that our model has learned how to resist this type of jailbreak attack, and our method only needs a small number of samples to fully learn how to defend. In addition, our other indicators remain stable during the test, which shows the robustness of our method.

\section{Algorithm of TIM}
\begin{algorithm}[h]
   \caption{The Pipeline of TIM}
   \label{alg:TIM}
\begin{algorithmic}

\STATE {\bfseries Initailize:} LLM ${\mathcal{E}_l,\mathcal{C}_d}$, Gist token $t_g$ and Detection Classifier $\mathcal{C}_d$, Jailbreak Memory $\mathcal{M}_j$, Detection Memory $\mathcal{M}_d$, Instruction Dataset $\mathcal{D}_{qa}$, Detection Dataset $\mathcal{D}_d$, Refusal Answer $t_{ref}$.
\STATE {\bfseries Input:} An instruction $t_{ins}$.
\STATE Generate the answer $t_{ans}$ of $t_{ins}$ by Equ. \eqref{equ:gen}
\STATE Obtain the jailbreak label by Equ. \eqref{equ:det1} and \eqref{equ:det2}.
\IF{jailbreak label equals to 1}
\STATE Append $(t_{ins},t_{ref})$ into $\mathcal{M}_j$.
\STATE Append $\{(t_{ins},t_{ref},0),(t_{ins},t_{ans},1)\}$ into $\mathcal{M}_d$.
\STATE Train the Adapter of $\mathcal{E}_l$ with $\mathcal{M}_j$ and $\mathcal{D}_{qa}$.
\STATE Train $t_g$ and $\mathcal{C}_d$ with $\mathcal{M}_d$ and $\mathcal{D}_d$
\ENDIF
\STATE {\bfseries Output:} Answer $t_{ans}$
\end{algorithmic}
\end{algorithm}
We summarize the pipeline of TIM in \Cref{alg:TIM}.

\vspace{20pt}
\section{Broader Impacts}
\label{sec:border_impact}
While this work does not directly target societal or community-level outcomes, it contributes to the broader scientific enterprise by advancing foundational understanding in jailbreak studies. The methods and findings presented may support future theoretical developments and inspire new directions in related research areas. Furthermore, the technical tools and insights generated can serve as a resource for researchers pursuing similar challenges, fostering further academic collaboration and exploration.

\section{Future Works and Limitations}
\label{sec:limit}
In practical applications, our method can be able to be combined with other static jailbreak attack defense methods to jointly improve the defense capabilities against jailbreak attacks. However, in this article, we did not verify the compatibility of TIM with other jailbreak attack defense methods. We plan to study this issue in subsequent work. In addition, due to the limitation of computing resources, we did not verify whether our method can be generalized to such models on a larger model (70 B+). In addition, our detector may have a decay when the detection token is extremely long. We consider adding data of different lengths to the detection dataset in future work to compensate for this limitation.

\end{document}